\documentclass[sigconf]{acmart}
\AtBeginDocument{%
  }

\setcopyright{acmlicensed}
\copyrightyear{2025}
\acmYear{2025}
\setcopyright{acmlicensed}\acmConference[KDD '25]{Proceedings of the 31st ACM SIGKDD Conference on Knowledge Discovery and Data Mining V.2}{August 3--7, 2025}{Toronto, ON, Canada}
\acmBooktitle{Proceedings of the 31st ACM SIGKDD Conference on Knowledge Discovery and Data Mining V.2 (KDD '25), August 3--7, 2025, Toronto, ON, Canada}
\acmDOI{10.1145/3711896.3736838}
\acmISBN{979-8-4007-1454-2/2025/08}

\usepackage{tabularx}
\usepackage{multirow}
\usepackage{makecell}
\usepackage{xcolor}

\newcommand{\red}[1]{\textcolor{black}{#1}}
\begin{document}

\title{AlphaAgent: LLM-Driven Alpha Mining with Regularized Exploration to Counteract Alpha Decay}

\author{Ziyi Tang}
\affiliation{%
  \institution{Sun Yat-sen University}
  \city{Guangzhou}
  \country{China}
}
\email{tangzy27@mail2.sysu.edu.cn}

\author{Zechuan Chen}
\affiliation{%
  \institution{Sun Yat-sen University}
  \city{Guangzhou}
  \country{China}
}
\email{chenzch6@mail2.sysu.edu.cn}

\author{Jiarui Yang}
\affiliation{%
  \institution{Sun Yat-sen University}
  \city{Guangzhou}
  \country{China}
}
\email{yangjr56@mail2.sysu.edu.cn}

\author{Jiayao Mai}
\affiliation{%
  \institution{University of New South Wales}
  \city{Sydney}
  \country{Australia}
}
\email{z5500756@ad.unsw.edu.au}

\author{Yongsen Zheng}
\affiliation{%
  \institution{Nanyang Technological University}
  \city{50 Nanyang Avenue}
  \country{Singapore}
}
\email{yongsen.zheng@ntu.edu.sg}

\author{Keze Wang}
\authornote{Corresponding author} 
\affiliation{%
  \institution{Sun Yat-sen University}
  \institution{Guangdong Key Laboratory of Big Data Analysis and Processing}
  \country{China}
  \institution{Peng Cheng Laboratory}
}
\email{kezewang@gmail.com}

\author{Jinrui Chen}
\affiliation{%
  \institution{The Chinese University of Hong Kong, Shenzhen}
  \city{Shenzhen}
  \country{China}
}
\email{120090765@link.cuhk.edu.cn}

\author{Liang Lin}
\affiliation{%
  \institution{Sun Yat-sen University}
  \institution{Guangdong Key Laboratory of Big Data Analysis and Processing}
  \country{China}
  \institution{Peng Cheng Laboratory}
}
\email{linliang@ieee.org}

\renewcommand{\shortauthors}{Ziyi Tang et al.}
\begin{abstract}

%

Alpha mining, a critical component in quantitative investment, focuses on discovering predictive signals for future asset returns in increasingly complex financial markets. However, the pervasive issue of alpha decay—where factors lose their predictive power over time—poses a significant challenge for alpha mining.
Traditional methods such as genetic programming are prone to rapid alpha decay, primarily due to their susceptibility to overfitting. At the same time, approaches driven by Large Language Models (LLMs), despite their promise, often fail to impose regularization against factor homogenization—resulting in crowded signals and accelerated decay. 
To address this challenge, we propose AlphaAgent, an autonomous framework that effectively integrates LLM-driven agents with ad hoc regularization for mining decay-resistant alpha factors. AlphaAgent employs three key mechanisms: (i) originality enforcement through a similarity measure based on abstract syntax trees (ASTs) against existing alphas, (ii) hypothesis—factor alignment via LLM-evaluated semantic consistency between market hypotheses and generated factors, and (iii) complexity control via AST-based structural constraints, preventing over-engineered constructions that are prone to overfitting. These mechanisms collectively guide the alpha generation process to balance originality, financial rationale, and adaptability to evolving market conditions, mitigating the risk of alpha decay. 
%
%
Extensive evaluations show that AlphaAgent outperforms traditional and LLM-based methods in mitigating alpha decay across bull and bear markets, consistently delivering significant alpha in Chinese CSI 500 and U.S. S\&P 500 markets over the past four years. Notably, AlphaAgent showcases remarkable resistance to alpha decay, elevating the potential for yielding powerful factors. Code is available at: \url{https://github.com/RndmVariableQ/AlphaAgent}

\end{abstract}




\maketitle


\section{Introduction}
\label{sec:intro}

Factor investing is a key strategy in modern quantitative finance, focusing on the systematic identification and exploitation of alpha factors (a.k.a., alphas) — quantifiable characteristics that can predict asset returns. However, alpha decay, or the decline of factor returns over time, arises from two main challenges. First, overfitting through excessive data mining ("p-hacking") leads to the emergence of spurious factors that appear significant in backtests but decay rapidly in real-world applications~\cite{fama1992cross}. Second, factor crowding occurs when too many investors adopt similar strategies, which can accelerate alpha decay and trigger sudden reversals during market stress~\cite{asness2013value,falck2022systematic,chen2024alpha}. This was evident in early 2024 with the size factor's underperformance in China's A-share market~\cite{wool2024china,CICC2024H2Outlook}, highlighting the risks of concentrated positioning in popular factors. Thus, it is rather crucial to counteract alpha decay during alpha mining.

%

Traditional methods for alpha mining mainly build on genetic programming (GP)~\cite{lin2019revisiting, 10.1145/1830483.1830584,zhaofan2022genetic,patil2023ai,alphagen,alphaevolve} and reinforcement learning (RL). However, they struggle to effectively address the alpha decay challenge. Traditional GP and RL approaches tend to over-emphasize the optimization of historical performance metrics while neglecting the underlying financial and economic rationale. Without sufficient consideration of financial soundness and economic intuition, these methods often produce alpha factors that show strong historical performance but experience rapid alpha decay when deployed in live markets. 
Large language models (LLMs) offer promising potential in alpha mining due to their extensive understanding of financial knowledge, which could help generate and evaluate factors beyond pure statistical significance. 
Despite their versatility, LLMs' direct application to alpha factor mining remains underutilized, revealing significant challenges in addressing alpha decay. The fundamental limitation lies in the lack of effective constraints against alpha decay in current LLM-based frameworks. \red{Without proper conceptual and implementational constraints to guide the mining process, LLMs' creativity to generate novel factors is significantly hampered~\cite{kumar2024largelanguagemodelsunlock,franceschelli2024creativity}. This leads to over-reliance on established factors (e.g., RSI~\cite{ctuaran2011relative} and momentum/value/size effects), exacerbating factor crowding—LLMs predominantly replicate existing market inefficiencies already exploited by participants. As a result, in rapidly evolving markets like the U.S. stock market, LLM-generated factors may struggle to uncover novel alpha}, leading to suboptimal investment performance.  

To address these issues, we propose a novel paradigm that effectively constrains LLM-based alpha mining to mitigate alpha decay, addressing key limitations of traditional approaches. At its core, AlphaAgent introduces three critical regularization mechanisms: (1) complexity control through symbolic expression trees and parameter counting, (2) hypothesis alignment via LLM-evaluated semantic consistency between market hypotheses and generated factors, and (3) novelty enforcement through a similarity measure based on abstract syntax trees (ASTs) against existing alpha libraries (e.g., Alpha101). 
AlphaAgent formalizes factor construction through an operator library and abstract syntax trees (ASTs), implementing a pairwise subtree isomorphism detection mechanism to quantify factor originality while using LLMs to verify financial intuition alignment through consistency scoring between hypotheses, descriptions, and expressions. These constraints guide LLMs to explore novel market inefficiencies while maintaining theoretical soundness, \red{alleviating the alpha decay challenge.} Based on these constraints, AlphaAgent builds an autonomous workflow encompassing hypothesis proposal, factor construction, factor development, backtesting, and feedback mechanisms. The framework begins with the \textit{idea agent} for the hypothesis proposal, with the first market insight provided by domain experts. Grounded in these hypotheses, the \textit{factor agent} constructs parsimonious and original factors to explore unexploited market inefficiencies, applying regularization mechanisms to balance complexity, novelty, and hypothesis alignment. The \textit{eval agent} then rigorously validates factors' executability and numerical stability while a backtesting system assesses their predictive effectiveness on historical data. \red{The feedback mechanism analyzes factor performance, guiding how to refine these factors in the next round.} This closed-loop process progressively derives a family of factors that capture emerging rather than overcrowded market inefficiencies, promoting alpha mining by balancing theoretical soundness with factor originality.

Extensive experiments demonstrate AlphaAgent's effectiveness in generating factors resistant to alpha decay. Through comprehensive evaluations in the Chinese CSI 500 and U.S. S\&P 500 markets from January 2021 to December 2024, our framework demonstrates stable factor performance across different market regimes. Through comprehensive evaluations in the Chinese CSI 500 and U.S. S\&P 500 markets from January 2021 to December 2024, our framework achieves an average annual excess return of 11.0\% (IR=1.5) and 8.74\% (IR=1.05) respectively after accounting for transaction costs, while maintaining robust performance across different market regimes. While traditional factors exhibit substantial decay in predictive power due to market crowding and overfitting, AlphaAgent's alphas maintain stable predictive effectiveness, demonstrating stronger resistance to alpha decay. The framework also displays efficiency in mining alphas, achieving an 81\% higher effective factor ratio (hit ratio) while consuming 30\% fewer tokens. These results substantially outperform traditional approaches and existing LLM-based approaches~\cite{
deepseekai2025deepseekr1incentivizingreasoningcapability,openai2024openaio1card,chen2024datacentric} in terms of both performance and performance persistence. 

Our main contributions can be summarized as follows:

\begin{itemize}
    \item We propose a systematic regularization mechanism to counteract alpha decay, combining originality enforcement, hypothesis alignment, and complexity control, facilitating the exploration of original and theoretically grounded alphas. 
    \item We implement a closed-loop multi-agent framework that evolves alphas with three agents that iteratively perform hypothesis generation, factor construction, and evaluation. 
    \item Extensive experiments reveal AlphaAgent's superior performance and resilience against alpha decay, attaining an 81\% improvement in hit ratio.
\end{itemize}

\section{Related Work}
Alpha factors, or mathematical expressions designed to predict future asset returns, have been a central focus in quantitative finance since Fama and French's pioneering work on their three-factor model~\cite{fama1992cross}. Traditional methods for alpha mining primarily rely on genetic programming (GP) for exploring the vast search space of factor formulation~\cite{lin2019revisiting, 10.1145/1830483.1830584,zhaofan2022genetic,patil2023ai}, yet they struggle to generate factors resistant to alpha decay. 
AlphaEvolve~\cite{alphaevolve} enriches traditional GP by incorporating parameter learning and matrix operations while maintaining GP's explicit formula structure. Shen et al.~\cite{shen2023mining} strengthen GP with sparsity constraints during mutation, which guide the search toward alpha factors with lower complexity. 
Another branch of traditional alpha factor mining relies on reinforcement learning (RL) to optimize factor formulation through policy learning~\cite{alphagen,shi2024alphaforgeframeworkdynamicallycombine,finrl}. AlphaGen~\cite{alphagen} mines formulaic alphas with deep reinforcement learning, using combination model performance as a reward signal to guide exploration within the alpha factor search space. Shi et al.~\cite{shi2024alphaforgeframeworkdynamicallycombine} propose an RL-based framework that simultaneously discovers and combines multiple alpha factors with fixed weights to form a unified signal. 
Building upon PPO~\cite{ppo}, an RL-based approach~\cite{zhao2024quantfactorreinforceminingsteady} is implemented that discards the critic network and introduces a reward-shaping mechanism aiming to generate more profitable and stable alphas for quantitative investment. 
\red{
However, these traditional paradigms face significant challenges in addressing alpha decay. Both the GP and RL approaches tend to overemphasize historical performance optimization, leading to either overly complex, overfitted factors or factors lacking rationale~\cite{shen2023mining, shi2024alphaforgeframeworkdynamicallycombine, zhao2024quantfactorreinforceminingsteady}. These limitations result in rapid alpha decay when factors are deployed in live markets.}

Recent advances in large language models (LLMs) offer a promising direction to address the alpha decay challenge, as LLMs excel at capturing evolving patterns and incorporating domain knowledge to generate more resilient factors~\cite{wang2023aligninglargelanguagemodels, haluptzok2023language, weng2023agent, sumers2024cognitivearchitectureslanguageagents}. While numerous studies have explored alpha mining using LLMs, the critical issue of alpha decay has received limited attention. AutoAlpha~\cite{zhang2020autoalphaefficienthierarchicalevolutionary} introduces an adaptive factor generation framework that continuously evolves alphas based on recent market conditions. LLMFactor~\cite{wang2024llmfactorextractingprofitablefactors} leverages knowledge-guided prompting to extract economically interpretable factors from financial news and historical data. FAMA~\cite{li-etal-2024-large-language} further advances this direction by dynamic factor combination and cross-sample selection, enabling performance across different market regimes. RD-Agent~\cite{yang2024collaborative} proposes a data-centric feedback loop that allows continuous factor adaptation to changing market conditions. However, existing approaches still suffer from alpha decay as they lack effective regularization mechanisms to prevent factors from overly relying on historical patterns and existing market knowledge.

\begin{figure*}[h]
\center
\includegraphics[width=0.85\textwidth]{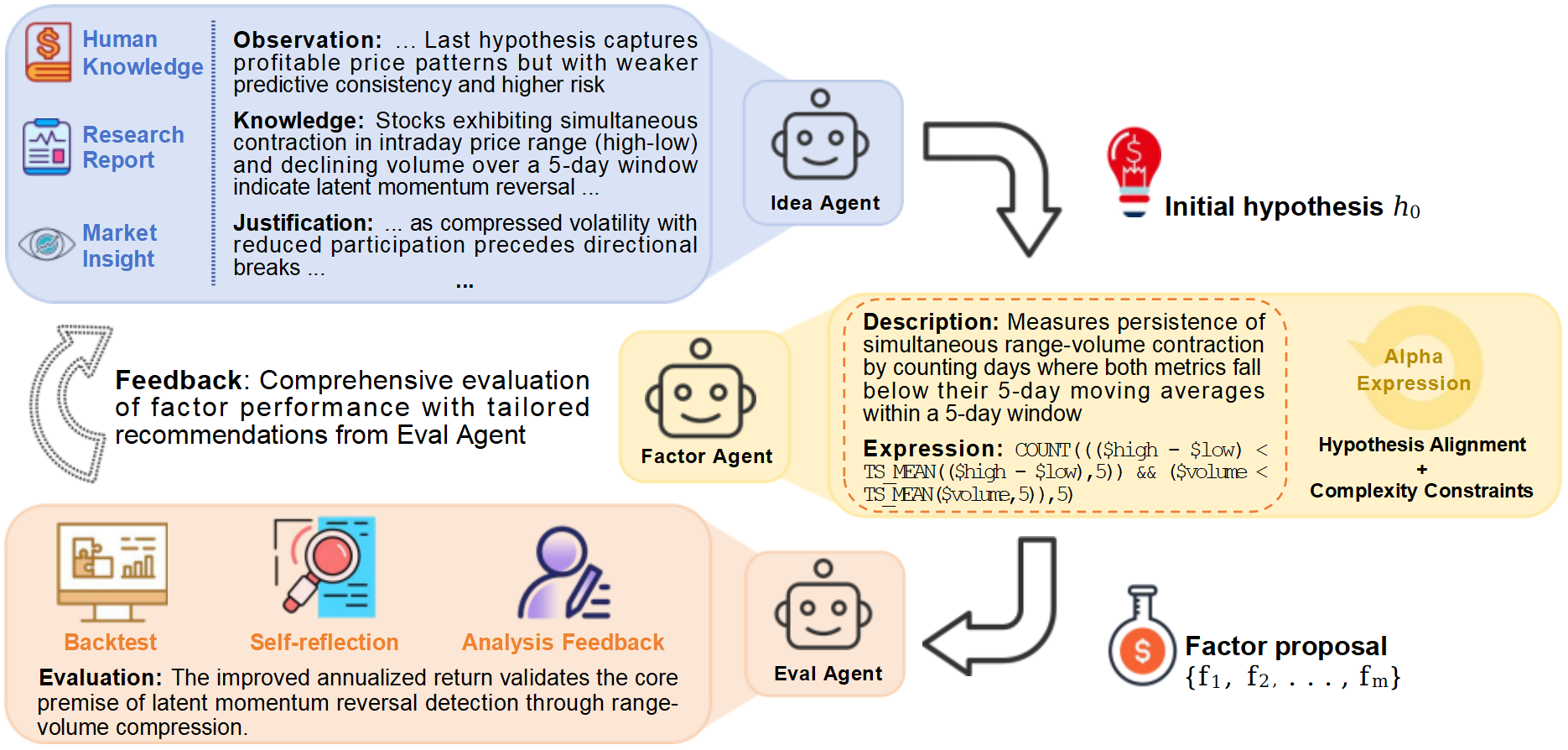}
\caption{The autonomous workflow of AlphaAgent, where three agents work collectively to mine alphas that balance financial rationale, originality, and adaptability to evolving market conditions, counteract the risk of alpha decay in alpha mining tasks.}
\label{fig:workflow}
\vspace{-10pt}
\end{figure*}

\section{AlphaAgent}
\subsection{Problem Formulation}
\label{subsec:problem_formulation}

The alpha mining task considers a set of stocks $\mathcal{S} = \{s_1, \dots, s_N\}$, a time window $\mathcal{T} = \{t_1, \dots, t_T\}$, and a feature matrix $\mathbf{X} \in \mathbb{R}^{N \times T \times D}$, where $D$ denotes the dimension of the raw features. The objective of alpha mining is to learn an alpha factor (or alpha) $f$ that maps a slice of input features $\mathbf{X}_{t}$ to a predictive signal $r_{t+1}$, namely the subsequent return. Formally, an alpha can be written as $f(\mathbf{X}_{t}) \rightarrow r_{t+1}$, where $r_{t+1}$ is the return on the day $t+1$. The alpha factor mining problem can be formulated as an optimization task:
\begin{equation}
\label{eq:baseline_optimization}
f^* = \arg\max_{f \in \mathcal{F}} \mathcal{L}\bigl(f(\mathbf{X}), \mathbf{y}\bigr) \;-\; \lambda\,\mathcal{R}(f),
\end{equation}
where $\mathcal{F}$ denotes the space of all possible factor expressions, $\mathbf{y}$ represents the ground-truth future returns (e.g., next-day returns), $\mathcal{L}$ measures predictive effectiveness (such as the information ratio or other performance metrics), $\mathcal{R}$ is a regularization term encouraging simplicity or novelty of the factor expression, and $\lambda$ is a balancing parameter that trades off between performance and complexity.

Distinct from conventional pure data-driven methods, we propose to leverage large language models (LLMs) as intelligent agents for alpha factor generation. Traditional methods often struggle to incorporate domain expertise effectively or tend to generate factors that lack economic intuition. LLMs, with their strong natural language understanding and reasoning capabilities~\cite{causalgpt,wang2023aligninglargelanguagemodels}, offer a promising solution by being able to comprehend and operationalize human market insights. \red{However, LLMs are inherently intractable due to their stochastic nature and limited capacity for instruction following}, which could lead to inconsistent or irrelevant alphas. To address these limitations, we introduce two critical aspects in the regularization term to ensure both the practical relevance and long-term effectiveness of the generated factors. Specifically, we introduce market hypotheses $h \in \mathcal{H}$ to guide the LLM-based factor construction process with domain-relevant insights (e.g., candlestick patterns, fundamental analysis results, market microstructure theories), and we ensure the generated factors maintain sufficient novelty. Concretely, we reformulate the above objective as:

\vspace{-5pt}
\begin{equation}
\label{eq:guided_optimization}
f^* = \arg\max_{f \in \mathcal{F}} \;\;\mathcal{L}\bigl(f(\mathbf{X}), \mathbf{y}\bigr) \;-\; \lambda\,\mathcal{R}_{g}(f, h),
\end{equation}

where the regularization term $\mathcal{R}_{g}(f, h)$ encompasses three components: (1) the complexity of the factor expression, (2) the alignment between the factor and a market hypothesis $h \in \mathcal{H}$ grounded in external knowledge (e.g. from domain experts), and (3) the novelty of the generated factor relative to existing ones. By incorporating these aspects into the regularization term, we ensure that the LLM-generated factors not only maintain theoretical relevance and practical interpretability through market hypotheses but also exhibit sufficient novelty to potentially capture unexploited market inefficiencies. Given the non-convex nature of the optimization objective, we employ an alternating optimization strategy between $\mathcal{L}$ and $\mathcal{R}_{g}$ to ensure convergence to a local minimum. The alternating procedure continues until a locally optimal alpha is found that balances predictive performance with the regularization constraints. This objective provides a flexible mechanism to balance predictive ability, domain soundness, and factor uniqueness, ultimately contributing to the long-term effectiveness of alpha factors. The detailed formulation of $\mathcal{R}_{g}(f, h)$ is delineated in Section \ref{subsec:factor_generation}.

\subsection{Factor Generation Modeling}
\label{subsec:factor_generation}
To operationalize the objective function defined in Eq. \ref{eq:guided_optimization}, a factor implementation mechanism is required that ensures both robustness and alignment with domain hypotheses. However, the inconsistent quality of LLM-generated outputs poses significant challenges in code-based factor construction. Approaches that generate code-based factors may frequently encounter operational barriers such as data format incompatibilities, inconsistencies across package versions, and difficulties in maintaining semantic coherence in extended code implementations. These challenges create a fundamental tension between code executability and semantic consistency, requiring LLMs to constantly balance these competing objectives in factor generation.

\subsubsection{Factor Parsing with Abstract Syntax Trees}

To address these limitations, we introduce an \emph{operator library} $\mathcal{O}$ that abstracts and standardizes various mathematical and financial operations (e.g., rolling minima/maxima, moving averages, conditional checks). This abstraction layer significantly streamlines the factor construction process by providing LLMs with a consistent and well-defined set of operations, thereby simplifying the semantic alignment between operator compositions and market hypotheses. The Operator Library serves as an intermediate representation that bridges the gap between high-level market insights and low-level implementation details, enabling more robust and maintainable factor generation. We define a \emph{parsing} procedure: 

\vspace{-5pt}
\begin{equation}
\label{eq:hypo_parsing}
\mathcal{G}: \bigl(\mathcal{H}, \mathcal{X}\bigr) \;\rightarrow\; \mathcal{F},
\end{equation}

\noindent where $\mathcal{H}$ represents the space of market hypotheses, with each hypothesis often described in a semi-structured form (refer to Sec. \ref{subsec:overview}). $\mathcal{X}$ denotes the space of raw features. The output space $\mathcal{F}$ contains tree-structured factors through \emph{symbolic assembly} from atomic operators in $\mathcal{O}$, where \emph{symbolic assembly} defines the process of composing and binding operators into factor expressions. By referring to $\mathcal{X}$, each factor can be bound to real data fields such as \verb|$price|, \verb|$volume|, and derivatives thereof.

We parse the textual hypothesis $h$ to a factor $f \in \mathcal{F}$ as an abstract syntax tree (AST), denoted as $T(f)$, via the following steps: (1) Identify key phrases in $h$ (e.g., \emph{triangle pattern''}, \emph{breakout''}) and map them to relevant operators in $\mathcal{O}$; (2) Assign numeric parameters (e.g., window size, threshold) for each operator based on $h$ or default domain values; (3) Assemble the operators into an abstract syntax tree $T(f)$, that captures the computational dependencies and execution flow of the factor expression. In $T(f)$, leaf nodes correspond to raw feature references (e.g., \verb|$high|, \verb|$low|), internal nodes represent operator instances (e.g., \verb|TS_MIN(.)|, \verb|SMA(.)|), and edges indicate the data flow between operations.

\subsubsection{Interpretability and Complexity Control}

Although objective \eqref{eq:baseline_optimization} focuses on maximizing predictive quality subject to domain alignment, it is equally crucial to control the complexity of any candidate expression. We incorporate a regularization term $\mathcal{R}_{g}(f)$ that penalizes overly complex syntax trees or large numbers of free hyperparameters. This ensures that the final solution not only adheres to the economic rationale encoded in $h$ but also remains interpretable and robust. For instance, we may define

\begin{equation}
\label{eq:regularization}
\begin{aligned}
\mathcal{R}_{g}(f, h) \;=\;\alpha_1 \cdot \mathrm{SL}(f) \;+\;\alpha_2 \cdot \mathrm{PC}(f) \;+\; \alpha_3 \cdot \mathrm{ER}(f, h),
\end{aligned}
\end{equation}

\noindent where $\mathrm{SL}(f)$ measures symbolic length, $\mathrm{PC}(f)$ counts free parameters (e.g., window lengths), and $ER(f,h)$ captures both the factor's novelty compared to existing alpha factors and its alignment with the given market hypothesis $h$. By tuning the weighting coefficients $\{\alpha_1,\alpha_2,\alpha_3\}$, we can obtain parsimonious yet powerful factor specifications.

\begin{figure}[!t]
\center
\includegraphics[width=0.49\textwidth]{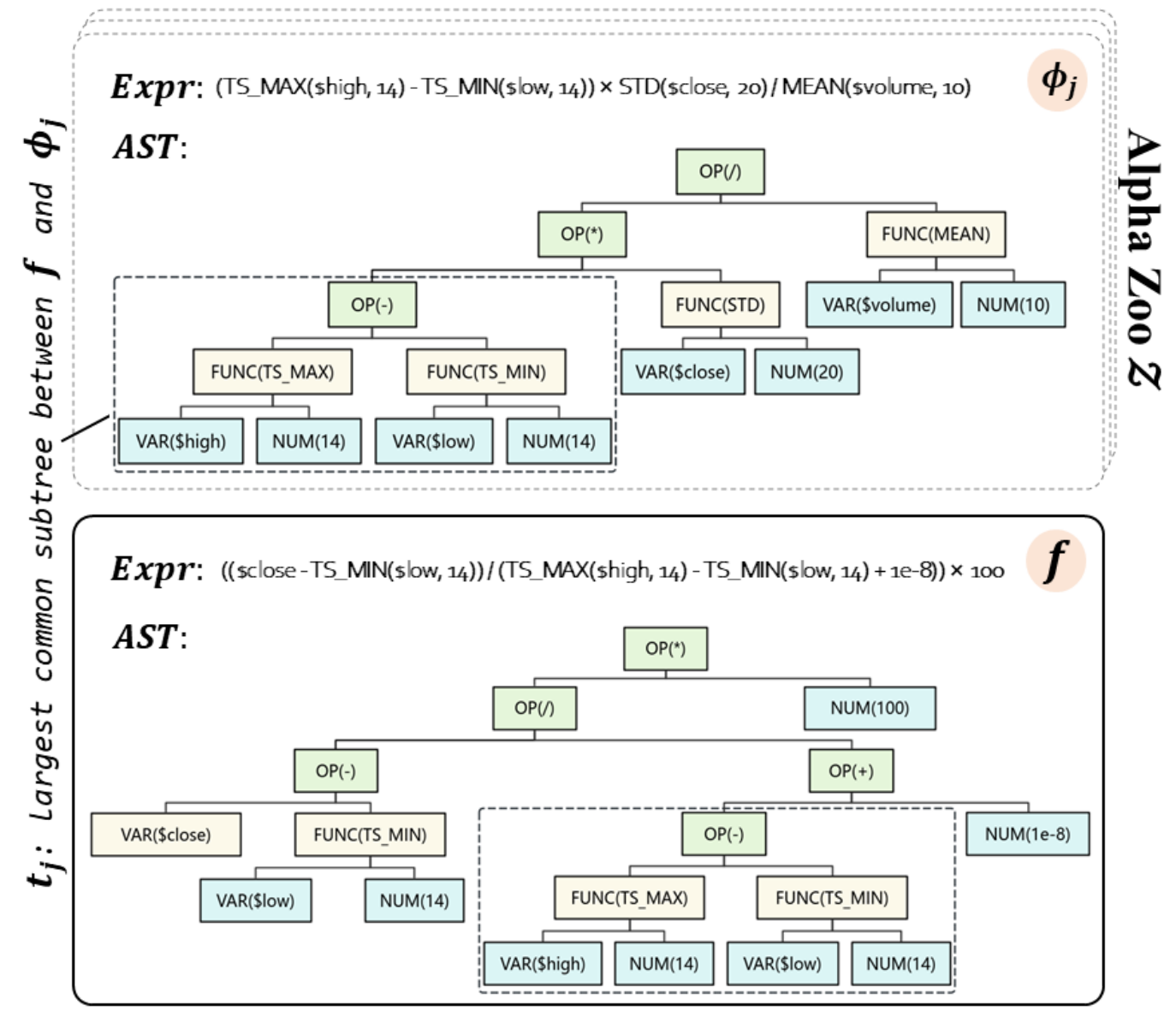}
\vspace{-10pt}
\caption{This figure shows the factor $f$ and an alpha zoo $\mathcal{Z}$ each represented as an expression or AST whose leaf nodes are depicted as light blue. The factor’s originality score is calculated by the maximum size of common subtrees between its AST and each factor within the alpha zoo. } 
\label{fig:ast}
\vspace{-12pt}
\end{figure}

To quantitatively assess the originality of proposed alpha factors and detect potential duplicates, we introduce a pair-wise factor similarity metric based on AST matching. For any given factor $f_i$, we first parse its expression into an AST representation $T(f_i)$. To compute the similarity between two ASTs $f_i$ and $f_j$, we identify their largest common subtree by recursively comparing their AST structures $T(f_i)$ and $T(f_j)$. The similarity metric $s$ is calculated as:

\begin{equation}
\label{eq:ast_pairwise_sim}
\begin{aligned}
s(f_i, f_j) = \max_{t_i \subseteq T(f_i), t_j \subseteq T(f_j)} \{|t_i| : t_i \cong t_j\},
\end{aligned}
\end{equation}

\noindent where $t_i$ and $t_j$ are subtrees of $T(f_i)$ and $T(f_j)$ respectively, $|t_i|$ denotes the size of the subtree (number of nodes), and $t_i \cong t_j$ indicates structural isomorphism between subtrees. With this similarity metric, a newly proposed factor can be compared with existing alphas that have been widely validated (see Fig. ~\ref{fig:ast}), such as Alpha101~\cite{alpha101}. Formally, we compute its maximum similarity score against an existing alpha zoo $\mathcal{Z} = \{\phi_1, \phi_2, ..., \phi_N\}$, providing a quantitative measure of the factor's originality, written as:

\begin{equation}
\label{eq:ast_ori}
\begin{aligned}
S(f) = \max_{\phi \in \mathcal{Z}} s(f, \phi),
\end{aligned}
\end{equation}

To ensure the semantic consistency between market hypotheses and generated factors, we employ LLMs to evaluate two critical alignments: (1) whether the factor description aligns with the market hypothesis as a valid implementation, and (2) whether the factor expression accurately reflects its description. For a given hypothesis $h$, factor description $d$, and factor expression $f$, we formulate a consistency scoring function:

\begin{equation}
\label{eq:alignment_score}
\begin{aligned}
\mathcal{C}(h, d, f) = \alpha c_1(h, d) + (1-\alpha) c_2(d, f),
\end{aligned}
\end{equation}

\noindent where $c_1(h, d) \in [0,1]$ evaluates whether the factor description $d$ represents a valid implementation of hypothesis $h$, $c_2(d, f) \in [0, 1]$ measures the consistency between the factor description and its mathematical expression, and $\alpha$ is a weighting parameter set to 0.5. For example, if a factor claims to capture market liquidity dynamics in its description but its expression contains no liquidity-related components (such as trading volume, bid-ask spread, or market depth), it would receive a low $c_2$ score, indicating potential misalignment between the claimed economic intuition and actual implementation. This scoring mechanism helps filter out factors that either deviate from the original market insight or contain mismatches between their semantic meaning and mathematical formulation, thereby reducing the risk of spurious factor generation.

Based on the above similarity metric and consistency evaluation, we can now formulate $\mathrm{ER}(f, h)$ that quantifies both the originality and hypothesis alignment of a generated factor:

\vspace{-5pt}
\begin{equation}
\label{eq:er_score}
\begin{aligned}
\mathrm{ER}(f, h) = & \beta_1 \cdot S(f) + \beta_2 \cdot \mathcal{C}(h, d, f) + \beta_3 \cdot \log(1 + |\mathcal{F}_f|) \\ 
\end{aligned}
\end{equation}

\noindent where $\beta_1, \beta_2, \beta_3$ are weighting coefficients, $\mathcal{F}_f$ represents the set of raw features used in factor $f$'s expression, and the logarithmic term penalizes excessive feature usage to promote factor parsimony. A lower $\mathrm{ER}$ score indicates better factor quality, with the first term penalizing similarity to existing factors, the second term ensuring hypothesis alignment, and the third term controlling expression complexity. By normalizing $S(f, \phi)$ to $[0,1]$ through appropriate scaling, all terms in the equation become comparable in magnitude.

Through the above factor generation process, each candidate $f$ is guaranteed to satisfy the originality, hypothesis alignment, and complexity constraints, while optional refinement heuristics can iteratively enhance or simplify expressions. The overall framework, described in subsequent sections, integrates \textit{eval agent} for factor evaluation, gathers feedback through reflective analysis, and ultimately establishes an autonomous framework to continuously refine and uncover alpha factors that counteract alpha decay.

\subsection{Autonomous Multi-Agent Framework}
\label{subsec:overview}
As illustrated in Figure ~\ref{fig:workflow}, AlphaAgent implements a recurrent framework for alpha factor mining through three specialized agents powered by large language models (LLMs): \textit{idea agent}, \textit{factor agent}, and \textit{eval agent}. The \textit{idea agent} synthesizes market hypotheses $h$ by integrating human knowledge, research reports, and market insights. Each hypothesis captures a potential market inefficiency pattern, such as value-momentum dynamics to behavioral biases and market structure anomalies~\cite{asness2013value,alpha101}. Next, the \textit{factor agent} translates these hypotheses into factor expressions that capture their underlying market dynamics. For each hypothesis $h$, the \textit{factor agent} generates multiple candidate implementations to quantify different aspects of the hypothesized inefficiency, including a natural language description of the factor logic and corresponding mathematical expressions using structural operators. The \textit{eval agent} evaluates them through backtesting on historical data, in-depth search for similar existing factors, and analysis of performance metrics. Performance feedback from the \textit{eval agent} guides the next iteration of hypothesis refinement and factor construction, forming a closed loop for continuous alpha mining. 

\paragraph{Idea Agent}
The \textit{idea agent} serves as the foundation of our framework by formalizing market hypotheses through a structured knowledge integration process. Drawing from external knowledge, the \textit{idea agent} employs the chain-of-thought reasoning~\cite{wei2022chain,causalgpt} to generate a market hypothesis with a systematic structure encompassing four interconnected components: \red{(1) \textit{observations} that provide empirical grounding through analysis of current market conditions or experimental results of previous rounds; (2) \textit{knowledge} that synthesizes established financial theories (e.g., market efficiency, behavioral finance), empirical market intuitions (e.g., momentum, mean reversion), and practitioners' conjectures derived from trading experience; (3) \textit{justification} that establishes theoretical soundness by linking observed patterns to underlying economic mechanisms; (4) \textit{specification} outlines implementation constraints, such as optional numeric or time-window parameters (e.g., ``10-day high/low''). In the initialization phase, the \textit{idea agent} generates the seed hypothesis $h_0$ based on a user-assigned research direction or market insight. The iterative evolving process leverages the initial hypothesis $h_0$ as an evolving anchor, where subsequent hypothesis generation is driven by feedback mechanisms informed through analysis of historical evolving traces. Through this structured approach, the \textit{idea agent} constructs market hypotheses that unite theoretical rigor with empirical validity, as a starting point for each round of alpha mining.}


\paragraph{Factor Agent}

The \textit{factor agent} serves as the bridge between theoretical market hypotheses and their quantitative manifestations, crafting alpha factor implementations through the regulated process outlined in Section \ref{subsec:factor_generation}. To enhance factor quality, the agent maintains an evolving knowledge base of both successful and failed factor implementations. Failed cases are categorized based on their failure modes, such as hypothesis misalignment and structural complexity violations. These failure cases are then encoded into the agent's knowledge base, allowing it to proactively avoid similar pitfalls in subsequent iterations. The agent employs a multi-stage refinement pipeline: first generating multiple candidate implementations for each hypothesis, then applying filters based on complexity and alignment metrics. During generation, the agent optimizes a group of alpha factors by referencing similar historical cases from its knowledge base, until they satisfy the originality, hypothesis alignment, and complexity constraints. This experiential learning mechanism enables the agent to continuously improve its generation capabilities, producing a variety of original, well-aligned, and parsimonious factor implementations over time.

\paragraph{Eval Agent}

The \textit{eval agent} first conducts a multi-dimensional evaluation of generated factors through a backtesting system. The evaluation process encompasses three primary aspects: predictive capability metrics that measure the factor's forecasting effectiveness, return performance metrics that assess the factor's profit-generating ability, and risk control metrics that evaluate its stability and robustness under various market conditions. Beyond quantitative assessment, the agent maintains an evaluation history to track factors' performance and identify emerging patterns in both successes and failures. This accumulated evaluation knowledge is systematically fed back to the \textit{idea agent}, enabling its hypothesis refinement. In this sense, the evaluation process not only validates individual factors but also continuously provides insights for the next round, forming a closed-loop mechanism to continuously optimize the overall alpha mining.

\section{Experiments}

\subsection{Experiment Settings}
\subsubsection{Metrics}
This study focuses on evaluating how AlphaAgent counteracts the alpha decay challenge against established baseline approaches using a comprehensive set of financial metrics. The information coefficient (IC) and rank information coefficient (RankIC) measure forecasting precision through the correlation between \red{daily predicted scores and actual daily returns, with IC using raw values and RankIC using ranked values.} Risk-adjusted performance indicators include the information coefficient information ratio (ICIR), which evaluates the consistency of IC performance by comparing its mean to standard deviation, and the information ratio (IR), which measures risk-adjusted excess returns relative to a benchmark. For absolute performance assessment, we use the annualized return (AR) to quantify the yearly excess investment return, and the maximum drawdown (MDD) to capture the largest peak-to-trough decline in portfolio value. \red{With these multi-faceted evaluation metrics, we can assess both the factor's long-term effectiveness and its resistance to alpha decay, as persistent IC/RankIC values and stable ICIR indicate sustained predictive power, while AR and MDD metrics reveal the practical impact of any deterioration in the factor's effectiveness over time. }

\subsubsection{Backtest Settings}
The backtesting experiments were conducted using the Qlib framework~\cite{yang2020qlibaiorientedquantitativeinvestment}, on CSI 500 of the Chinese A-share market and S\&P 500 of the U.S. stock market, spanning 2021 to 2024. Raw data employed to construct alpha factors include only OHLCV (i.e., \$open, \$high, \$low, \$close, \$volume). CSI 500 data is collected from Baostock~\cite{baostock}, while S\&P 500 data is from Yahoo Finance~\cite{yfinance}. See Table ~\ref{tab:dataset_splits} for precise dataset splits.

\begin{table}[h!]
\centering
\small
\renewcommand{\arraystretch}{1.15}  
\begin{tabular}{lcc}
\toprule
\textbf{Asset} & \textbf{Period} & \textbf{Trading Days} \\
\midrule
        & Training: 2015-01 to 2019-12 & 1258 \\
S\&P 500 & Validation: 2020-01 to 2020-12 & 253 \\
        & Testing: 2021-01 to 2025-01 & 1004 \\
\hline
        & Training: 2015-01 to 2019-12 & 1219 \\
CSI 500  & Validation: 2020-01 to 2020-12 & 243 \\
        & Testing: 2021-01 to 2025-01 & 968 \\
\bottomrule
\end{tabular}
\vspace{3pt}
\caption{Periods of Training, validation, and testing splits, with their trading days for S\&P 500 and CSI 500.}
\label{tab:dataset_splits}
\end{table}
\vspace{-10pt}

AlphaAgent employs GPT-3.5-turbo~\cite{gpt35} as the foundational LLM to support agents. For RD-Agent~\cite{chen2024datacentric}, following its authors, GPT-4-turbo~\cite{gpt4} is used. Four fundamental alphas, including \textit{intra-day return}, \textit{daily return}, \textit{20-day relative volume}, and \textit{normalized daily range}, serve as the base alphas and will be concatenated with newly proposed alphas to train a LightGBM model~\cite{NIPS2017_6449f44a}. Before being fed into LightGBM, both features and returns undergo cross-sectional Z-score normalization to ensure comparability across stocks. The LightGBM model, with a maximum depth of 4 layers, is responsible for forecasting the next-day returns. In backtesting, a top-$k$ dropout strategy is employed to select the 50 top-ranked stocks based on the predicted returns and exclude the 5 lowest-ranked stocks. All backtesting results account for transaction fees. For CSI 500, the transaction fees are set at 0.0005 for buying and 0.0015 for selling. For S\&P 500, only selling fees are applied, with a rate of 0.0005.

\subsubsection{Baselines}

We compare AlphaAgent with several baseline approaches: (1) Traditional time-series forecasting models including \textbf{LSTM}~\cite{graves2012long} and \textbf{Transformer}~\cite{10.5555/3295222.3295349} which capture temporal dependencies; (2) Tree-based model \textbf{LightGBM}~\cite{NIPS2017_6449f44a} for handling structured financial data; (3) Specialized financial models including \textbf{StockMixer}~\cite{Fan_Shen_2024} and \textbf{TRA}~\cite{HengxuKDD2021}, which focus on integrating multiple trading strategies and handling non-i.i.d. market patterns respectively; (4) Agent-based approaches including \textbf{AlphaForge}~\cite{shi2024alphaforgeframeworkdynamicallycombine} and \textbf{RD-Agent}~\cite{chen2024datacentric}, which leverage deep learning and LLMs for alpha factor generation and optimization; (5) Deep reasoning models, \textbf{OpenAI-\textit{o1}}~\cite{openai2024openaio1card} and \textbf{DeepSeek-\textit{R1}}~\cite{deepseekai2025deepseekr1incentivizingreasoningcapability}.

\begin{table*}[ht!]
  \footnotesize
  \captionsetup{justification=centering}
  \caption{Performance Comparison of Different Methods on CSI 500 (China) and S\&P 500 (U.S.). The AR shown in this table denotes the annualized excess return. \textbf{Bold} and \underline{underlined} numbers represent the best and second-best performance across all compared approaches, respectively. }
  \vspace{-3pt}
  \label{tab:performance}
  \begin{tabular}{lcccccccccc}
    \toprule
    \textbf{Market} & \multicolumn{5}{c}{\textbf{CSI 500} (2021-01-01 to 2024-12-31)} & \multicolumn{5}{c}{\textbf{S\&P 500} (2021-01-01 to 2024-12-31)} \\
    \cmidrule(lr){2-6} \cmidrule(lr){7-11}
    \textbf{Method} & IC & ICIR & AR & IR & MDD & IC & ICIR & AR & IR & MDD \\
    \midrule
    LSTM~\cite{graves2012long} & 0.0175 & 0.1521 & \underline{4.96\%} & \underline{0.6225} & \underline{-9.68\%} & 0.0028 & 0.0181 & -1.51\% & -0.1671 & -26.05\% \\
    Transformer~\cite{10.5555/3295222.3295349} & 0.0131 & 0.1074 & 4.11\% & 0.5074 & -17.45\% & 0.0013 & 0.0129 & -4.55\% &  -0.4964 & -34.96\% \\
    LightGBM~\cite{NIPS2017_6449f44a} & 0.0120 & 0.1209 & -1.18\% & -0.1588 & -18.97\% & 0.0011 & 0.0116 & -2.64\% & -0.4224 &  -21.17\%  \\
    TRA~\cite{HengxuKDD2021} & \underline{0.0198} & \underline{0.1794} & 2.91\% & 0.4261 & -12.73\% & -0.0003 & -0.0027 & -8.51\% & -1.1345 & -49.55\% \\
    Stock-Mixer~\cite{Fan_Shen_2024} & 0.0000 & 0.0003 & -0.35\% & -0.0496 & -16.82\% & 0.0030 & 0.0312 & -2.49\% & -0.3342 & -29.43\% \\
    AlphaForge~\cite{shi2024alphaforgeframeworkdynamicallycombine} & 0.0146 & 0.1299 & 3.45\% & 0.3270 & -17.67\% & 0.0026 & 0.0240 & 2.45\% & \underline{0.3369} & \underline{-10.91\%} \\
    RD-Agent~\cite{chen2024datacentric} & 0.0113 & 0.0872 & 0.78\% & 0.0744 & -20.85\% & 0.0019 & 0.0123 & 1.69\% & 0.1664 & -23.18\% \\
    DeepSeek-\textit{R1}~\cite{deepseekai2025deepseekr1incentivizingreasoningcapability} \textit{best-of-10} & 0.0132 & 0.1201 & 1.58\% & 0.2086 & -14.95\% & \underline{0.0048} & \underline{0.0369} & \underline{2.75\%} & 0.2400 & -15.34\%  \\
    OpenAI-\textit{o1}~\cite{openai2024openaio1card} \textit{best-of-10} & 0.0159 & 0.1502 & 0.46\% & 0.0632 & -21.29\% & 0.0028 & 0.0217 & 2.29\% & 0.2021 & -16.35\% \\
    AlphaAgent & \textbf{0.0212} & \textbf{0.1938} & \textbf{11.00\%} & \textbf{1.488} & \textbf{-9.36\%} & \textbf{0.0056} & \textbf{0.0552} & \textbf{8.74\%} & \textbf{1.0545} & \textbf{-9.10\%} \\
    \bottomrule
  \end{tabular}
\end{table*}

\begin{figure*}[ht]
\center
\includegraphics[width=0.9\textwidth]{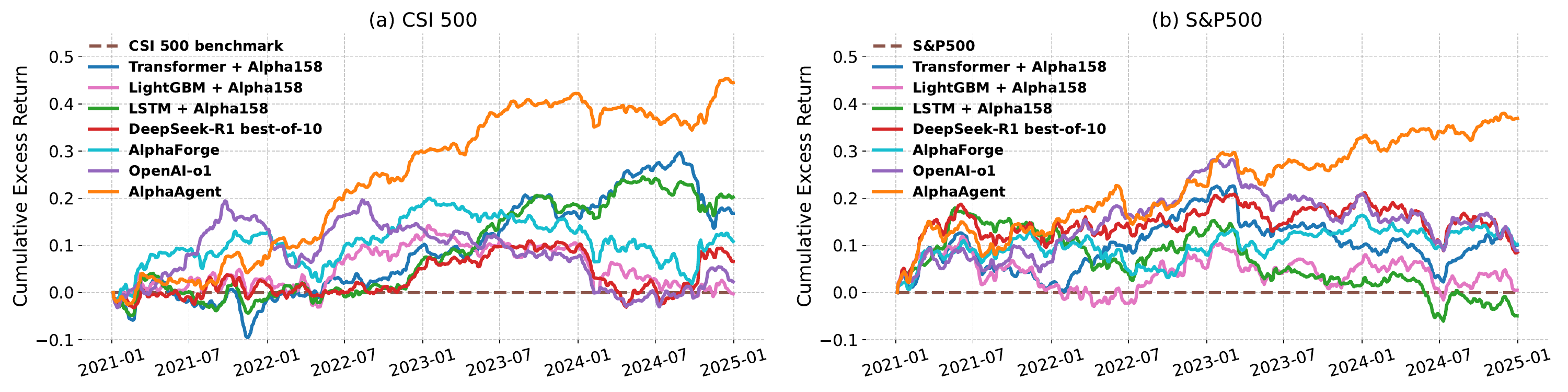}
\vspace{-10pt}
\caption{Cumulative excess returns of different approaches on CSI 500 and S\&P 500.} 
\label{fig:ex_return}
\vspace{-10pt}
\end{figure*}

\subsection{Overall Performance}
In Table \ref{tab:performance}, we show a comparison of the overall performance of the different methods for four years, from January 1, 2021, to December 31, 2024, in the CSI 500 (China) and S\&P 500 (U.S.) stock markets. Regarding AlphaForge, RD-Agent, Deepseek-R1, OpenAI-o1, and AlphaAgent, we apply their respective output alphas directly in this comparative analysis. For reasoning models Deepseek-R1 and OpenAI-o1, we streamline the task prompt of AlphaAgent to facilitate alpha mining and evolution across five iterative rounds, with each round providing corresponding backtesting results using the identical configuration of AlphaAgent. For RD-Agent and AlphaAgent, we conduct 20 independent trials, with each trial comprising five evolutionary rounds. In each trial, we insert the optimal factors into their respective alpha zoos, yielding the optimal combination as the final result. The performance of each method is evaluated by five key metrics: IC, ICIR, AR, IR, and MDD. Bold numbers indicate the best-performing methodology in each dataset, while underlined numbers represent sub-optimal performance.

Over the four-year period from January 2021 to January 2025, AlphaAgent consistently outperformed other models across all key metrics in both markets. For predictive power, it achieved the highest IC (0.0212) and ICIR (0.1938) in CSI 500 and similarly led in S\&P 500 with IC of 0.0056 and ICIR of 0.0552. In terms of returns, AlphaAgent generated the best annualized returns of 11.00\% in CSI 500 and 8.74\% in S\&P 500, significantly ahead of the second-best performers (LSTM's 4.96\% and Deepseek-R1's 2.75\% respectively). The model also demonstrated superior risk management, maintaining MDDs below 10\% in both markets (-9.36\% in CSI 500 and -9.10\% in S\&P 500), while achieving the highest risk-adjusted returns with IRs of 1.488 and 1.0545, respectively.

Figure \ref{fig:ex_return} (a) and (b) illustrate the cumulative excess returns of different models in the CSI 500 and S\&P 500 markets from 2021 to early 2025, revealing distinct patterns of alpha decay across different models and market environments. The time series models (LSTM and Transformer) exhibit second-tier performance in the CSI 500 market with no significant excess returns before 2023 while suffering severe decay in the S\&P 500 market with consistently negative returns since 2023-02, particularly evident in Transformer's notable drawdown after 2024-08. In contrast, AlphaAgent demonstrates remarkable resilience in both markets, maintaining relatively stable quarterly performance with approximately 45\% cumulative excess returns in CSI 500 and exceeding 37\% in S\&P 500 throughout the testing period. Non-sequential models like LightGBM with Alpha158 show complete alpha decay in the S\&P 500 market as their cumulative excess returns oscillate around zero, while DeepSeek-R1's performance declines after 2023 despite its strong reasoning capabilities, suggesting a lack of systematic constraints. This stark contrast in alpha persistence between models and markets indicates that maintaining consistent alpha generation is notably more challenging in the more efficient U.S. market environment, though AlphaAgent remains robust in such conditions. 


\subsection{Alpha Decay Analysis}
\label{subsec:alpha_decay}

\begin{figure}[h!]
\center
\includegraphics[width=0.45\textwidth]{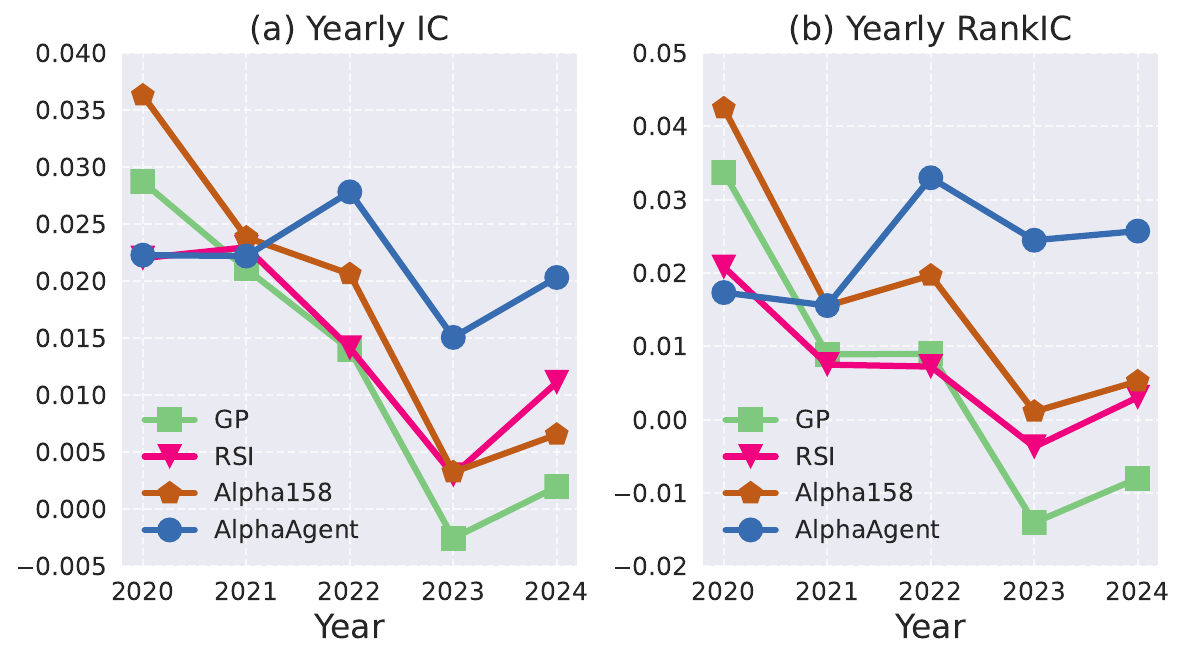}
\vspace{-10pt}
\caption{Yearly IC and RankIC comparison on CSI 500. While other factors' predictive power rapidly decays over time, 15 alphas mined by AlphaAgent maintain stable performance.} 
\label{fig:ic_ric_CSI 500}
\vspace{-10pt}
\end{figure}

Figure \ref{fig:ic_ric_CSI 500} compares the yearly performance between Alpha158~\cite{yang2020qlibaiorientedquantitativeinvestment}, GP~\cite{lin2019revisiting}, a technical indicator RSI, and AlphaAgent's alphas over 5 years on CSI 500. Alpha158, GP, and RSI all exhibit substantial declines in predictive power, with their ICs dropping from 0.022-0.036 to near zero and RankICs decreasing from 0.020-0.042 to around zero, highlighting the widespread challenge of alpha decay in the Chinese stock market. In contrast, AlphaAgent's alphas demonstrate remarkable stability, maintaining predictive effectiveness with IC values consistently around 0.02 and RankIC values around 0.025 throughout the period. This contrast highlights the superior sustainability of AlphaAgent compared to traditional factors, which exhibit stronger signs of alpha decay.


This analysis reveals fundamental differences between how traditional alphas and our approach face modern financial market challenges. GP's rapid performance deterioration suggests potential overfitting to historical patterns, making it particularly vulnerable to changing market conditions. Meanwhile, the significant performance degradation of Alpha158 and RSI Indicator exemplifies the "alpha decay" phenomenon described in Sec. \ref{sec:intro}, to a great extent, stemming from market crowding as these strategies become widely adopted by investors. When multiple market participants simultaneously execute similar trading strategies based on the same factors, their collective actions can diminish the factors' predictive power. In contrast, AlphaAgent's sustained performance suggests that its factor modeling mechanism works to explore sustained and less exploited alphas, effectively mitigating the crowding effect that plagues traditional approaches. 

\subsection{Alpha Mining Efficiency Analysis}
\label{subsec:vs_rdagent}

\begin{figure}[!h]
\center
\includegraphics[width=0.4\textwidth]{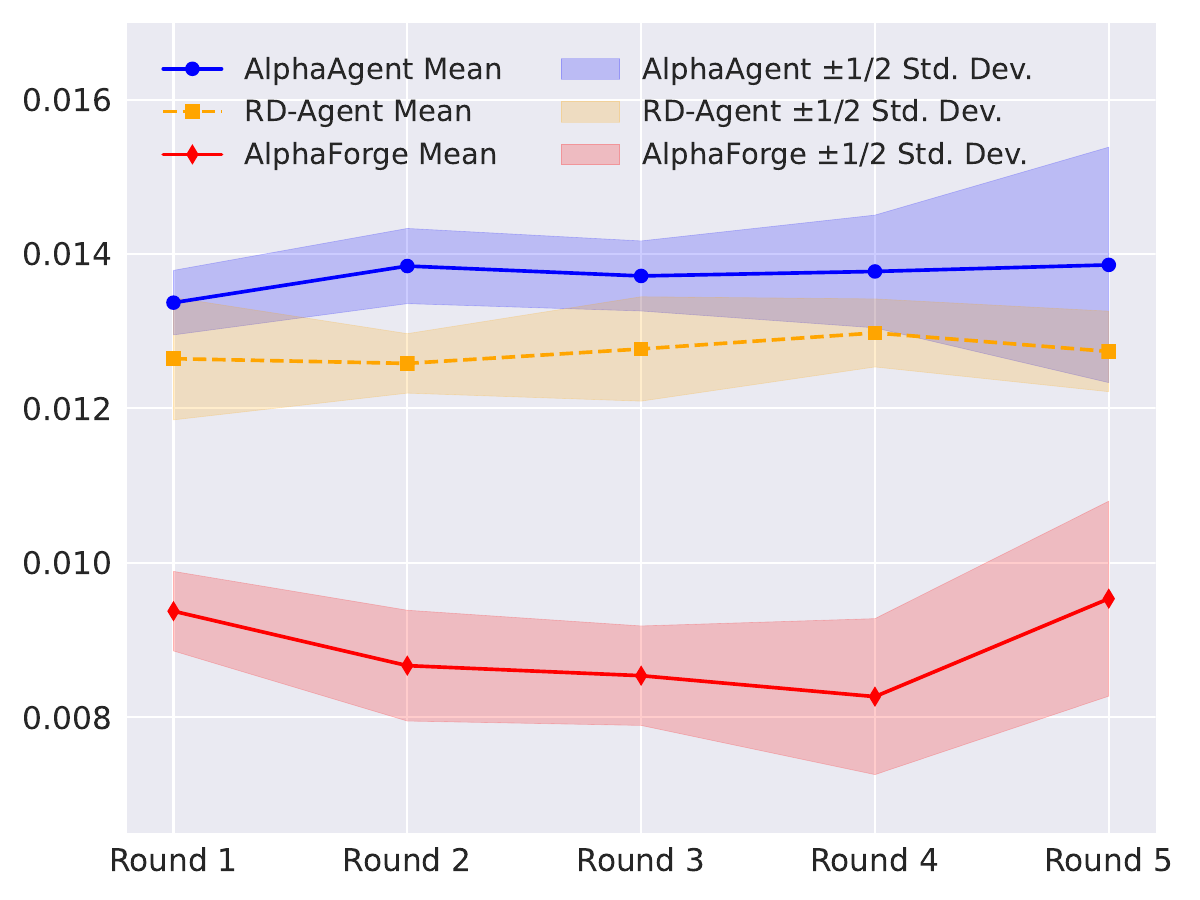}
\vspace{-10pt}
\caption{The evolution of the ICs in the first five rounds. } 
\label{fig:ic_std}
\vspace{-5pt}
\end{figure}

This subsection conducts an analysis that focuses on the quality of generated alpha factors and the computational efficiency of the generation process. Figure~\ref{fig:ic_std} illustrates the evolution of IC values across five rounds for AlphaAgent and its counterparts on CSI 500's test split. First, the results show that RD-Agent exhibits relatively stable and smaller variance (shown by the consistent width of its shaded region), likely due to its lack of exploration incentives for LLM-based agents, indicating more homogeneous factor candidates. Such candidates may lead to factor crowding and accelerated alpha decay as similar signals become widely exploited in the market. On the contrary, AlphaAgent consistently maintains higher average IC values compared to RD-Agent as well as AlphaForge throughout all rounds, demonstrating the superior predictive power of its generated factors gained from complexity and hypothesis-alignment constraints against overfitting and financial rationality of the generated factors. A notable observation is an increasing variance (shown by the expanding shaded region) of AlphaAgent's IC values as the rounds progress, suggesting AlphaAgent's factor diversity and potentially a wider exploration space brought by its originality penalties, which could lead to a higher probability of discovering effective factors. AlphaAgent's originality penalties drive broad exploration of the factor space, while these complementary mechanisms ultimately contribute to consistently higher average IC values.

\subsection{Ablation Study}

\begin{figure}[!h]
\center
\includegraphics[width=0.48\textwidth]{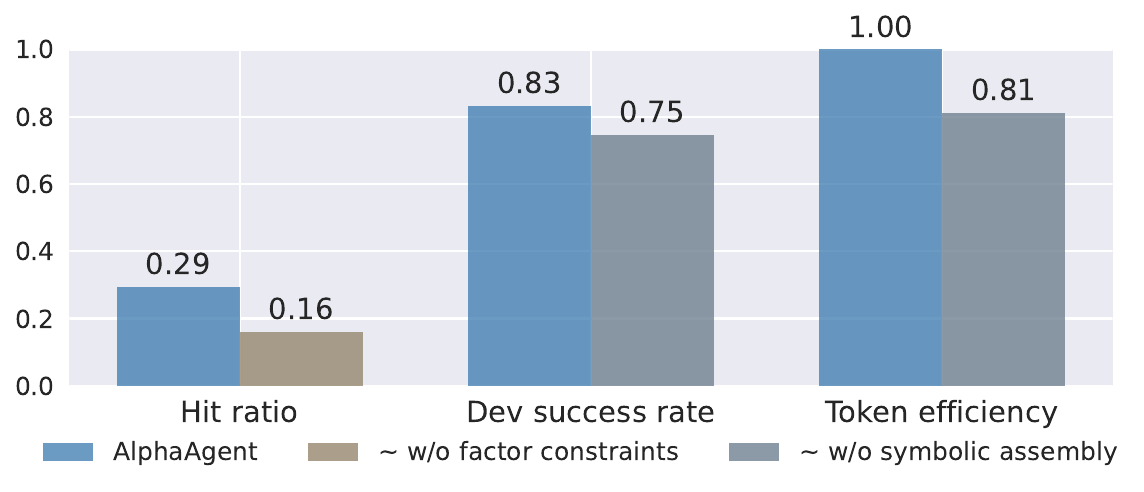}
\vspace{-15pt}
\caption{A comparison of RD-Agent and AlphaAgent regarding hit ratio, development success rate, and token efficiency.} 
\label{fig:eff}
\vspace{-5pt}
\end{figure}

In Figure ~\ref{fig:eff}, we ablate AlphaAgent's core components, i.e., factor modeling constraints and symbolic assembly, across three key metrics. This evaluation is based on 100 rounds of evolution, evenly split between the CSI 500 and S\&P 500 markets. The hit ratio measures the proportion of generated alphas achieving exceptional returns per round (>4.0\% annualized for CSI 500 and >1.8\% for S\&P 500, representing the top 5\% of all generated alphas). The dev success rate captures the percentage of factors that can be successfully executed without any code defects or numerical errors. \red{Token efficiency measures the inverse ratio of average tokens consumed per candidate factor generation, with the higher efficiency normalized to 1.0 while maintaining relative proportions. }

\begin{figure*}[h!]
\center
\includegraphics[width=0.85\textwidth]{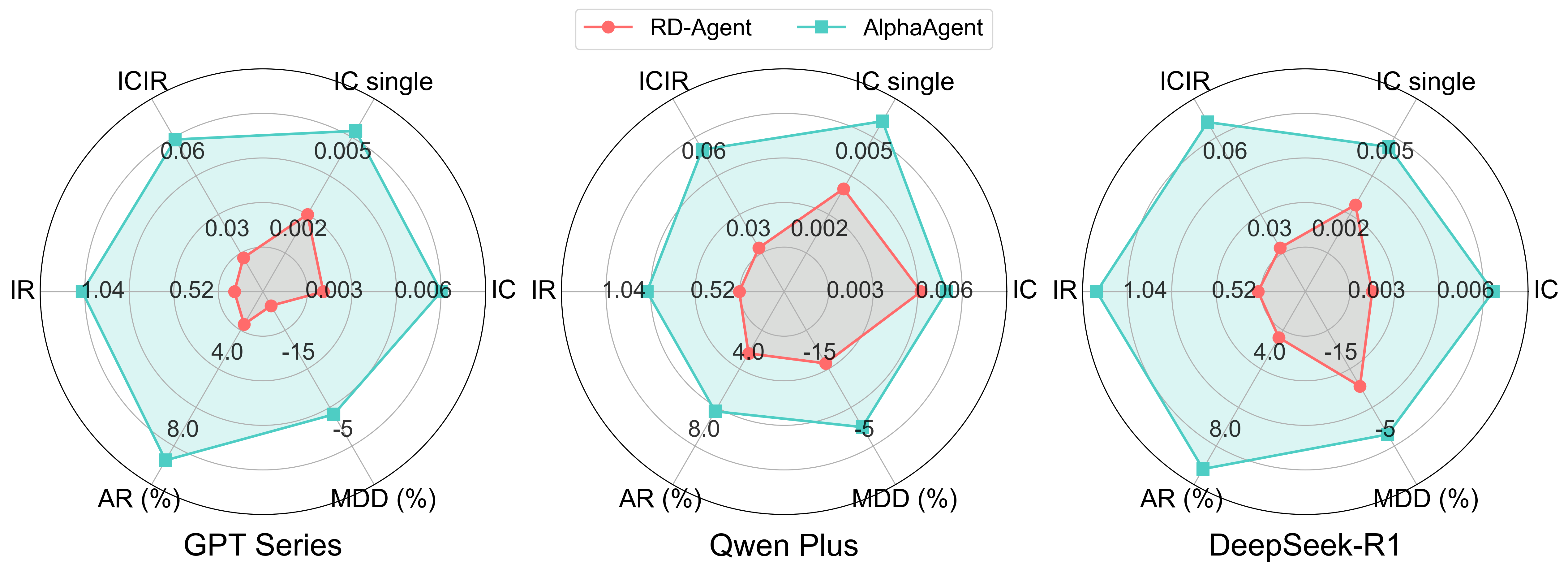}
\vspace{-10pt} 
\caption{AlphaAgent's Performance with different base LLMs on S\&P 500 from 2021 to 2024. IC$_{\textit{single}}$ represents the highest IC value among individual factors. The annual standard deviation (annual std.) is computed based on the yearly average IC values across four years.} 
\label{fig:llms}
\vspace{-10pt}
\end{figure*}

Our ablation studies demonstrate AlphaAgent's effectiveness across these three metrics. In terms of hit ratio, AlphaAgent achieves 0.29 compared to 0.16 without factor modeling constraints, representing an 81\% improvement. This substantial increase indicates that incorporating constraints significantly enhances the quality of generated alpha factors, more likely to mine sustained and predictive alphas. The development success rate comparison (0.83 vs 0.75) evaluates the impact of symbolic assembly—\red{it streamlines factor implementation by reducing potential code defects and numerical errors.} When examining token efficiency, AlphaAgent achieves higher efficiency, demonstrating a 23\% improvement in generating viable factor candidates per token. These results collectively validate that our proposed mechanisms effectively improve factor generation across multiple dimensions while relaxing computational overhead.


\red{To evaluate the performance of AlphaAgent across different base LLMs, we conduct a comparative analysis of GPT-3.5-turbo~\cite{gpt35}, Qwen-Plus~\cite{qwen2025qwen25technicalreport}, and DeepSeek-R1~\cite{deepseekai2025deepseekr1incentivizingreasoningcapability} on the S\&P 500 dataset, with the results illustrated in Fig. \ref{fig:llms}. On one hand, these radar charts demonstrate a consistent improvement in factor quality with stronger models. Particularly, as the only reasoning LLM, DeepSeek-R1 achieves the highest ICIR (0.0615) and annualized return (9.19\%) with the lowest drawdown (-6.50\%), outperforming GPT-3.5-turbo and Qwen-Plus. On the other hand, while comparing with RD-Agent across these LLMs, \textbf{Student's t test~\cite{rice2007mathematical} confirms the statistical significance of AlphaAgent's improvements}, with p-values for IC differences between AlphaAgent and RD-Agent being consistently below 0.05 (GPT-3.5-turbo: 0.0311, Qwen-Plus: 0.0109, DeepSeek-R1: 0.0382). These results collectively demonstrate that while the choice of base LLM significantly impacts factor quality (with DeepSeek-R1 delivering optimal performance), AlphaAgent provides statistically significant improvements over its counterpart across all model variants, validating its advancement over existing approaches.}

\subsection{Implications}
The results shown in Sec. \ref{subsec:alpha_decay} and Sec. \ref{subsec:vs_rdagent} underscore a critical insight into modern quantitative finance: the imperative for alpha mining methods to possess continuous exploration capability beyond fitting historical patterns and relying on established theories. This is particularly evident in today's highly efficient markets where traditional statistical arbitrage strategies face diminishing returns due to increased competition and market adaptation. Our experiments demonstrate that approaches incorporating active exploration mechanisms can better identify and capitalize on emerging market inefficiencies before they dissipate in rapidly evolving markets. With the emergence of LLMs and their growing application in financial analysis, the landscape of alpha mining is undergoing fundamental changes, requiring approaches to be more adaptive and exploratory than ever before. These findings highlight the need for future quantitative trading research to focus on advanced exploration frameworks that balance the exploitation of known patterns with the exploration of novel alpha sources, particularly in an era where AI-driven market analysis is becoming increasingly prevalent.

\section{Conclusion}
This paper introduces AlphaAgent, a novel LLM-driven framework that effectively counteracts the critical challenge of alpha decay with three key regularization mechanisms: originality enforcement, complexity control, and hypothesis alignment. By incorporating these mechanisms into an autonomous framework, AlphaAgent produces decay-resistant and performant alpha factors while maintaining theoretical soundness, achieving substantial excess returns with remarkable consistency and decay resistance through various market conditions. The framework also suggests a promising direction for the next-generation alpha mining framework that swiftly adapts to market evolution while maintaining alpha sustainability.


\section{Acknowledgment}
This work was supported in part by the National Natural Science Foundation of China (NSFC) under Grant 62276283, in part by the China Meteorological Administration's Science and Technology Project under Grant CMAJBGS202517, in part by Guangdong Basic and Applied Basic Research Foundation under Grant 2023A1515012985, in part by Guangdong-Hong Kong-Macao Greater Bay Area Meteorological Technology Collaborative Research Project under Grant GHMA2024Z04, and in part by Fundamental Research Funds for the Central Universities, Sun Yat-sen University under Grant 23hytd006.

\newpage
\bibliographystyle{ACM-Reference-Format}
\bibliography{reference}

\end{document}